\documentclass[preprint,prl,aps,letterpaper]{revtex4}
\usepackage{natbib}
\usepackage{graphicx}

\newcommand{\ppbar}{$p\overline{p}$~}

\newcommand{\ttbar}{$t\overline{t}$~}
\newcommand{\MET}{\mbox{$\raisebox{.3ex}{$\not$}E_T$\hspace*{0.5ex}}}

\newcommand{\intlum}{$\int\!{\cal L}dt$}

\def\simge{\mathrel{%
   \rlap{\raise 0.511ex \hbox{$>$}}{\lower 0.511ex \hbox{$\sim$}}}}
\def\simle{\mathrel{
   \rlap{\raise 0.511ex \hbox{$<$}}{\lower 0.511ex \hbox{$\sim$}}}}
\def\lessim{\mathrel {\vcenter {\baselineskip 0pt \kern 0pt
\hbox{$<$} \kern 0pt \hbox{$\sim$} }}}
\def\gessim{\mathrel {\vcenter {\baselineskip 0pt \kern 0pt
\hbox{$>$} \kern 0pt \hbox{$\sim$} }}}
\def \rightdownarrow
 {\kern.4em {\raise1.75ex\hbox{$\Bigl |$}$\kern-0.27em{\longrightarrow}$}}
\def \mrightdownarrow
 {\kern.4em {\raise1.25ex \hbox{$|$} \kern-0.27em{
                                                \longrightarrow}}}

\def\GeVc{GeV$\!/c$}

\def\mMEt{\not\kern-.35em {E_T}}
\def\MEt{\hbox{$\mMEt$}}

\font\eightit=cmti8
\def\r#1{\ignorespaces $^{#1}$}
\begin{document}
\begin{large}
\begin{center}
Measurement of the $t\bar{t}$ Production Cross Section in 
$p\bar{p}$ Collisions at $\sqrt{s} = 1.96$ TeV using Dilepton Events
\end{center}
\end{large}
 
\font\eightit=cmti8
\def\r#1{\ignorespaces $^{#1}$}
\hfilneg
\begin{sloppypar}
\noindent
D.~Acosta,\r {15} T.~Affolder,\r 8 T.~Akimoto,\r {53}
M.G.~Albrow,\r {14} D.~Ambrose,\r {42} S.~Amerio,\r {41}  
D.~Amidei,\r {32} A.~Anastassov,\r {49} K.~Anikeev,\r {30} A.~Annovi,\r {43}
% ...
\end{sloppypar}
\vskip .026in
\begin{center}
(CDF Collaboration)
\end{center}

\vskip .026in
\begin{center}
\r 1  {\eightit Institute of Physics, Academia Sinica, Taipei, Taiwan 11529, 
Republic of China} \\
\r 2  {\eightit Argonne National Laboratory, Argonne, Illinois 60439} \\
\r 3  {\eightit Istituto Nazionale di Fisica Nucleare, University of Bologna,
I-40127 Bologna, Italy} \\
\r 4  {\eightit Brandeis University, Waltham, Massachusetts 02254} \\
\r 5  {\eightit University of California at Davis, Davis, California  95616} \\
\r 6  {\eightit University of California at Los Angeles, Los 
Angeles, California  90024} \\
\r 7  {\eightit University of California at San Diego, La Jolla, California  92093} \\ 
\r 8  {\eightit University of California at Santa Barbara, Santa Barbara, California 
93106} \\ 
\r 9 {\eightit Instituto de Fisica de Cantabria, CSIC-University of Cantabria, 
39005 Santander, Spain} \\
\r {10} {\eightit Carnegie Mellon University, Pittsburgh, PA  15213} \\
\r {11} {\eightit Enrico Fermi Institute, University of Chicago, Chicago, 
Illinois 60637} \\
\r {12}  {\eightit Joint Institute for Nuclear Research, RU-141980 Dubna, Russia}
\\
\r {13} {\eightit Duke University, Durham, North Carolina  27708} \\
\r {14} {\eightit Fermi National Accelerator Laboratory, Batavia, Illinois 
60510} \\
\r {15} {\eightit University of Florida, Gainesville, Florida  32611} \\
\r {16} {\eightit Laboratori Nazionali di Frascati, Istituto Nazionale di Fisica
               Nucleare, I-00044 Frascati, Italy} \\
\r {17} {\eightit University of Geneva, CH-1211 Geneva 4, Switzerland} \\
\r {18} {\eightit Glasgow University, Glasgow G12 8QQ, United Kingdom}\\
\r {19} {\eightit Harvard University, Cambridge, Massachusetts 02138} \\
\r {20} {\eightit The Helsinki Group: Helsinki Institute of Physics; and Division of
High Energy Physics, Department of Physical Sciences, University of Helsinki, FIN-00044, Helsinki, Finland}\\
\r {21} {\eightit Hiroshima University, Higashi-Hiroshima 724, Japan} \\
\r {22} {\eightit University of Illinois, Urbana, Illinois 61801} \\
\r {23} {\eightit The Johns Hopkins University, Baltimore, Maryland 21218} \\
\r {24} {\eightit Institut f\"{u}r Experimentelle Kernphysik, 
Universit\"{a}t Karlsruhe, 76128 Karlsruhe, Germany} \\
\r {25} {\eightit High Energy Accelerator Research Organization (KEK), Tsukuba, 
Ibaraki 305, Japan} \\
\r {26} {\eightit Center for High Energy Physics: Kyungpook National
University, Taegu 702-701; Seoul National University, Seoul 151-742; and
SungKyunKwan University, Suwon 440-746; Korea} \\
\r {27} {\eightit Ernest Orlando Lawrence Berkeley National Laboratory, 
Berkeley, California 94720} \\
\r {28} {\eightit University of Liverpool, Liverpool L69 7ZE, United Kingdom} \\
\r {29} {\eightit University College London, London WC1E 6BT, United Kingdom} \\
\r {30} {\eightit Massachusetts Institute of Technology, Cambridge,
Massachusetts  02139} \\   
\r {31} {\eightit Institute of Particle Physics, McGill University,
Montr\'{e}al, Canada H3A~2T8; and University of Toronto, Toronto, Canada
M5S~1A7} \\
\r {32} {\eightit University of Michigan, Ann Arbor, Michigan 48109} \\
\r {33} {\eightit Michigan State University, East Lansing, Michigan  48824} \\
\r {34} {\eightit Institution for Theoretical and Experimental Physics, ITEP,
Moscow 117259, Russia} \\
\r {35} {\eightit University of New Mexico, Albuquerque, New Mexico 87131} \\
\r {36} {\eightit Northwestern University, Evanston, Illinois  60208} \\
\r {37} {\eightit The Ohio State University, Columbus, Ohio  43210} \\  
\r {38} {\eightit Okayama University, Okayama 700-8530, Japan}\\  
\r {39} {\eightit Osaka City University, Osaka 588, Japan} \\
\r {40} {\eightit University of Oxford, Oxford OX1 3RH, United Kingdom} \\
\r {41} {\eightit University of Padova, Istituto Nazionale di Fisica 
          Nucleare, Sezione di Padova-Trento, I-35131 Padova, Italy} \\
\r {42} {\eightit University of Pennsylvania, Philadelphia, 
        Pennsylvania 19104} \\   
\r {43} {\eightit Istituto Nazionale di Fisica Nucleare, University and Scuola
               Normale Superiore of Pisa, I-56100 Pisa, Italy} \\
\r {44} {\eightit University of Pittsburgh, Pittsburgh, Pennsylvania 15260} \\
\r {45} {\eightit Purdue University, West Lafayette, Indiana 47907} \\
\r {46} {\eightit University of Rochester, Rochester, New York 14627} \\
\r {47} {\eightit The Rockefeller University, New York, New York 10021} \\
\r {48} {\eightit Istituto Nazionale di Fisica Nucleare, Sezione di Roma 1,
University di Roma ``La Sapienza," I-00185 Roma, Italy}\\
\r {49} {\eightit Rutgers University, Piscataway, New Jersey 08855} \\
\r {50} {\eightit Texas A\&M University, College Station, Texas 77843} \\
\r {51} {\eightit Texas Tech University, Lubbock, Texas 79409} \\
\r {52} {\eightit Istituto Nazionale di Fisica Nucleare, University of Trieste/\
Udine, Italy} \\
\r {53} {\eightit University of Tsukuba, Tsukuba, Ibaraki 305, Japan} \\
\r {54} {\eightit Tufts University, Medford, Massachusetts 02155} \\
\r {55} {\eightit Waseda University, Tokyo 169, Japan} \\
\r {56} {\eightit Wayne State University, Detroit, Michigan  48201} \\
\r {57} {\eightit University of Wisconsin, Madison, Wisconsin 53706} \\
\r {58} {\eightit Yale University, New Haven, Connecticut 06520} \\
\end{center}

\date{\today}

\begin{abstract}
We report a measurement of the \ttbar production cross section 
using dilepton events with jets and missing transverse
energy in \ppbar collisions at a center-of-mass energy of 1.96~TeV.
Using a 197 $\pm$ 12~pb$^{-1}$ data sample recorded by the upgraded Collider
Detector at Fermilab, we use two complementary techniques to select
candidate events.
We compare the number of observed events and selected kinematical distributions
with the predictions of the Standard Model and find good agreement.
The combined result of the two techniques yields 
a \ttbar production cross section of 
$7.0^{+2.4}_{-2.1}\mbox{(stat.)}^{+1.6}_{-1.1}\mbox{(syst.)}\pm 
0.4\mbox{(lum.)}$~pb.

\end{abstract}

\pacs{14.65.Ha, 13.85.Qk, 12.38.Qk}
\keywords{top quarks, leptons, quantum chromodynamics}

\maketitle

  Since the discovery of the top quark~\cite{discovery}, experimental attention
has turned to the examination of its production and decay properties.
Within the Standard Model (SM),
the top quark production cross section is calculated with an uncertainty of
$\simle 15$\%~\cite{topxsect1,topxsect2}. 
Furthermore, in the SM, 
the top quark decays to a $W$ boson
and $b$ quark $\sim 100$\% of the time.  The $W$ subsequently
decays to either a pair of quarks or a lepton-neutrino pair. 
Measuring the rate of the reaction 
$\mbox{\ppbar} \rightarrow \mbox{\ttbar} \rightarrow 
b\ell^+\nu_\ell\bar{b}\ell^{\prime -}\bar{\nu}_{\ell^\prime}$
tests both the production and decay mechanisms of the top quark.
A significant deviation from the SM prediction would indicate 
either a novel production mechanism, {\it e.g.} a heavy resonance 
decaying into \ttbar 
pairs~\cite{resonance}, or a novel decay mechanism, {\it e.g.} 
decay into supersymmetric particles~\cite{susydecay}.
The CDF and D\O\ collaborations previously measured 
the \ttbar production cross section in the dilepton
channel during Run~I of the Fermilab Tevatron~\cite{cdfd0top}.
These and
related measurements were consistent with SM expectations
but suffered large uncertainties due to small event samples.

   This Letter describes a measurement of the
\ttbar cross section in the dilepton channel using data
from Run~II of the Tevatron taken with the upgraded
Collider Detector at Fermilab (CDF~II).   The data sample
corresponds to an integrated luminosity of 197 $\pm$ 12~pb$^{-1}$~\cite{lum}, 
$\sim 2 \times$ that used in Run~I.   Moreover, we expect
the higher center-of-mass energy of 1.96~TeV in Run II 
to increase the production of \ttbar events by $\sim 30$\%
relative to the Run~I rate at 1.8~TeV~\cite{topxsect1,topxsect2}.  
The upgrades to the CDF~II detector further increase the \ttbar 
yield with improved lepton acceptance.
We perform two complementary analyses of the new data.
One, inspired by the technique used by CDF in Run I,
requires that both leptons be specifically identified as either
electrons or muons (``DIL'' analysis).  The other
technique allows one of the leptons  to be identified
only as a high-$p_T$, isolated track (``LTRK'' analysis), thereby 
significantly increasing the lepton detection efficiency 
with some increase in expected background events.

  The CDF~II detector~\cite{cdf2det} is an azimuthally and
forward-backward symmetric apparatus designed to 
study \ppbar~reactions at the Tevatron.  
The detector has a charged particle tracking system 
immersed in a 1.4~T magnetic field, aligned coaxially 
with the \ppbar beams.   A silicon microstrip detector
provides tracking over the radial range 1.5 to 28~cm.
A 3.1~m long open-cell drift chamber, the Central Outer Tracker (COT), 
covers the radial range from 40 to 137~cm. 
The fiducial region of the silicon detector extends to $|\eta|\sim 2$~\cite{geomnote}, while the
COT provides coverage for $|\eta|\simle 1$.
 
  Segmented electromagnetic and hadronic sampling calorimeters surround the 
tracking system and measure the energy flow of interacting particles in
the pseudo-rapidity range $|\eta|<3.6$. 
This analysis uses the new end-plug detectors to identify electron 
candidates with $1.2<|\eta|<2.0$ in addition to the central
detectors for lepton candidates with $|\eta|<1.1$. 
A set of drift chambers located outside the central hadron calorimeters 
and another set behind a 60~cm iron shield detect energy deposition from
muon candidates with $|\eta|\le0.6$. 
Additional drift chambers
and scintillation counters detect muons in the region $0.6\le|\eta|\le1.0$.
Gas Cherenkov counters located in the $3.7<|\eta|<4.7$ region~\cite{clc}
measure the average number of inelastic \ppbar collisions 
per bunch crossing and thereby determine the beam luminosity.

  The $b\ell^+\nu_\ell\bar{b}\ell^{\prime -}\bar{\nu}_{\ell^\prime}$ events
under study produce two high-$p_T$ leptons, missing transverse
energy (\MET)~\cite{geomnote} from the undetected neutrinos, and two jets from
the hadronization of the $b$ quarks.  Additional jets are
often produced by initial-state and final-state radiation.   A trigger
system first identifies candidate events by finding either a central
electron or muon candidate with $E_T>18$~GeV~\cite{etpt}, or an end-plug
electron candidate with $E_T>20$~GeV~\cite{geomnote} in an 
event with \MET$>15$~GeV.  After full event reconstruction, the candidate event sample 
is further refined by selection criteria determined {\it a priori} to minimize 
the expected statistical and systematic uncertainties of the cross section result.
  
  Both analyses require two oppositely charged leptons with $E_T>20$~GeV~\cite{etpt}.  
One lepton, the ``tight'' lepton, must pass strict lepton
identification requirements and be isolated. A lepton is 
isolated if the total $E_T$
within a cone $\Delta R\equiv\sqrt{(\Delta\eta)^2 + (\Delta\phi)^2} \le 0.4$,
minus the lepton $E_T$, is $<10$\% of the lepton $E_T$~\cite{etpt}.
Tight electrons have a well-measured track pointing at an energy deposition in
the calorimeter.  For electrons with $|\eta| > 1.2$, this track 
association uses a calorimeter-seeded silicon tracking 
algorithm~\cite{phoenix}.  In addition, the candidate's electromagnetic shower
profile must be consistent with that expected for electrons.  
Tight muons must have a well-measured track linked to
hits in the muon chambers and energy deposition in the calorimeters 
consistent with that expected for muons.

  The other lepton, the ``loose'' lepton, is identified
differently by the two analyses.  The DIL analysis requires the 
loose lepton to be an electron or muon selected as above, with the
exceptions that it need not be isolated and muon identification requirements are relaxed.  
The LTRK analysis defines a loose 
lepton as a well-measured, isolated track with $p_T>20$~\GeVc\ 
in the range of $|\eta|< 1$ where the isolation requirement is the tracking 
analog of the calorimetric isolation employed for tight leptons.  
These selections add acceptance for dilepton events
where electrons or muons pass through gaps in the calorimetry or muon systems. 
They also contribute acceptance for
single prong hadronic decays of the $\tau$ lepton from $W\rightarrow\tau\nu$.
Consequently, the LTRK analysis derives 20\% of its acceptance from taus, compared
with 12\% for the DIL analysis.



  Candidate events must have \MET $>25$~GeV.
To reduce the occurrence of false \MET due to mismeasured jets, we 
require that the \MET vector point away from any jet.
Each analysis takes additional steps
to further suppress false \MET arising from 
mismeasurement of their respective loose leptons. 
The DIL analysis requires that the \MET vector 
be at least $20^{\circ}$ from the closest lepton.
The LTRK analysis corrects the \MET
for all loose leptons whenever the 
associated calorimeter 
$E_T$ is $<70$\% of the
track $p_T$.  It further rejects events for which 
the \MET vector lies within $5^{\circ}$ of the loose lepton axis.
In both analyses, these additional topological cuts are not applied in
events with \MET $>50$~GeV. 

  The DIL (LTRK) analysis counts jets with $E_T>15~(20)$~GeV detected in
$|\eta|<2.5~(2.0)$,
where we define a jet as a fixed-cone cluster with a cone size of
$R=0.4$.  We correct jet $E_T$ measurements for the
effects of 
calorimeter non-uniformity and 
absolute energy scale~\cite{escale}.

  After removal of cosmic-ray muons and photon-conversion electrons,
the dominant backgrounds to dilepton \ttbar events are Drell-Yan 
($q\bar{q} \rightarrow Z/\gamma^{\star}\rightarrow \ell^+\ell^-$) production, 
``fake'' leptons in $W\rightarrow \ell\nu$ + jet events where a jet is falsely
reconstructed as a lepton candidate, and diboson ($WW$, $WZ$, and $ZZ$)
production.  
Drell-Yan events typically have little \MET.
Thus, for events with dilepton invariant mass within 15 GeV/$c^2$ of the $Z$ boson
resonance, the DIL analysis imposes a cut on the
ratio of \MET to the sum of the jet $E_T$'s projected along the \MET vector,
whereas the LTRK analysis tightens its \MET requirement to 
$\MEt > 40$~GeV. To estimate residual Drell-Yan sample contamination  
we utilize both a PYTHIA~\cite{pythia} Monte Carlo calculation
with detector simulation and the data itself.
We select $Z$ boson candidates in the mass
range of 76--106~GeV/$c^{2}$ and count the number of events passing nominal 
and Drell-Yan-specific selection criteria.  
After subtraction of expected non-Drell-Yan contributions, these two numbers provide
the normalization for the distribution of expected contributions inside and outside 
the $Z$ boson mass window.  This distribution is obtained 
as a function of jet multiplicity using a sample of simulated 
events.

  We estimate the fake lepton background contribution by
applying a fake lepton rate to a data sample of $W\rightarrow \ell \nu$~+~jet 
events.  We determine this fake rate using
a large sample of events triggered by at least one jet with $E_T>50$~GeV
after removing sources of real leptons such as $W$ and $Z$ decays.  
To check the accuracy and robustness of this estimate we apply our 
fake lepton rate to different samples with varied physics content:
jet data with 20, 70 and 100~GeV trigger thresholds, an inclusive photon
sample, and an inclusive electron sample.  
The observed numbers of fake leptons agree with our fake rate predictions
within statistical uncertainties ({\it e.g.} 74 observed vs. $70 \pm 14$ predicted for LTRK).
An additional check is performed on the like-sign subset of the dilepton sample itself, 
which is dominantly $W\rightarrow \ell \nu$~+~jet events with one fake lepton.
We compare the number of observed like-sign events to the like-sign
fake background predictions and find good agreement ({\it e.g.} 5 observed vs. $6.3 \pm 1.4$ predicted for DIL).  

  We determine geometric and kinematic acceptance for the diboson backgrounds  
using PYTHIA and ALPGEN+HERWIG Monte Carlo calculations~\cite{alpgen,herwig}\ 
followed by a simulation of the CDF~II detector.  We use the CTEQ5L
parton distribution functions~\cite{cteq5l}\ to model the
momentum distribution of the initial state partons.  
We normalize the total number of expected events for these processes to their theoretical 
cross-sections: $13.3$ pb for $WW$, $4.0$ pb for $WZ$ and
$1.5$ pb for $ZZ$~\cite{diboson}. We estimate the uncertainty 
in these background predictions by comparing different 
Monte Carlo calculations for the same diboson process.  
Similarly, we obtain the acceptance for $t\bar{t}$ 
using a PYTHIA Monte Carlo 
calculation assuming $m_{top}=175$~GeV/$c^2$ and
$BR(W\rightarrow \ell
\nu)=10.8\%$. 

  We present the predicted and observed numbers of oppositely charged
dilepton events versus jet multiplicity in Table~\ref{events}. 
Good agreement is seen for the background-dominated zero and one jet events, 
establishing confidence in the background estimates detailed above.
We measure the \ttbar production cross section 
using events with two or more jets. The DIL analysis
enhances its signal sensitivity by requiring that $H_T$, 
the scalar sum of the lepton $p_T$,  jet $E_T$, 
and \MET, be $>200$~GeV.  

\begin{table*}[t]
\begin{ruledtabular}
\begin{tabular}{l|ccc|cccc}
 &\multicolumn{3}{c|}{LTRK}&\multicolumn{4}{c}{DIL}\\
\hline
 &$N_{\mbox{jet}}=0$&$N_{\mbox{jet}}=1$&$N_{\mbox{jet}}\geq 2$
 &$N_{\mbox{jet}}=0$&$N_{\mbox{jet}}=1$&$N_{\mbox{jet}}\geq 2$
 &$H_T>200$~GeV \\
\hline
$WW,WZ,ZZ$
           & $21.8\pm 5.2$ & $6.3\pm 1.5$ &$1.2\pm 0.3$
           & $11.4\pm 3.3$ & $3.2\pm 0.9$ &$1.1\pm 0.3$
           & $0.7\pm 0.2$ \\
Drell-Yan 
          & $26.5\pm 9.8$ & $16.4\pm 6.0$ & $4.2\pm1.6$
          &  $4.4\pm 1.9$ & $2.9\pm 1.1$ &$1.3\pm 0.5$
          & $0.9\pm 0.5$\\
 Fakes 
          & $16.5\pm 2.4$ & $5.0\pm 1.0$ & $1.5\pm 0.5$
          &  $3.0\pm 1.2$ & $2.4\pm 1.0$ &$1.5\pm 0.6$
          & $1.1\pm 0.5$\\
\hline
Total Background 
          & $64.8\pm 11.3$ & $27.7\pm 6.3$ & $6.9\pm 1.7$
          &  $18.8\pm 4.0$ & $8.5\pm 1.8$ &$3.9\pm 0.9$
          & $2.7\pm 0.7$\\

\hline
Expected \ttbar 
          & $0.3\pm 0.2$ & $3.4\pm 0.6$ & $11.5\pm 1.5$
          &  $0.1\pm 0.0$ & $1.3\pm 0.2$ &$8.5\pm 1.2$
          & $8.2\pm 1.1$\\
Total  
          & $65.1\pm 11.3$ & $31.1\pm 6.3$ & $18.4\pm 2.3$
          &  $18.9\pm 4.0$ & $9.8\pm 1.9$ & $12.4\pm 1.6$
          & $10.9\pm1.4$\\
Observed
          & 73 & 26 & 19
          & 16 &  9 & 14
          & 13\\
\end{tabular}
\end{ruledtabular}
\caption{\label{events}Expected background and \ttbar contributions ($m_{top}=175$~GeV/$c^{2}$,
$\sigma$=6.7~pb) compared with observed data.}
\end{table*}

  Systematic uncertainties 
include uncertainties on the acceptance times efficiency
$(a \times \epsilon)$ and the background estimates.
The dominant uncertainty on $a \times \epsilon$
is due to uncertainties on the jet energy scale 
and lepton identification/isolation efficiencies.
The background uncertainty is dominated by the 
statistical uncertainty in the Drell-Yan contribution
arising from 
the limited number of $Z$ events with high \MET.
Table~\ref{unc} lists all systematic uncertainties.

\begin{table}[h]
\begin{tabular}{lcc}
\hline\hline
Signal and Background Uncertainties & LTRK & DIL\\
\hline
Lepton(track) ID &  5\%(6\%) & 5\%\\
Jet energy scale - signal & 6\% & 5\%\\
Jet energy scale - background & 10\% & 18--29\%\\
Initial/final state radiation & 7\% & 2\%\\
Parton distribution functions & 6\%  & 6\%\\
Monte Carlo Generators & 5\% & 6\% \\
$WW,WZ,ZZ$ estimate & 20\% & 20\% \\
Drell-Yan Estimate & 30\% & 51\% \\
Fake Estimate & 12\% & 41\% \\
\hline\hline
\end{tabular}
\caption{\label{unc}Summary of systematic uncertainties.}
\end{table}

Using Table~\ref{events}, 
the expected signal-to-background
ratios are 3.1 for the DIL analysis and 1.7 for the LTRK analysis. 
The products  $a \times \epsilon \times BR(t\bar{t} \rightarrow b\ell^+\nu_\ell\bar{b}\ell^{\prime -}\bar{\nu}_{\ell^\prime})$ are 
$(0.62\pm 0.09)$\% and $(0.88\pm 0.12)$\%, respectively.
The total integrated luminosity is 
\intlum=$197\pm 12$~pb$^{-1}$.
Hence, the measured cross sections, 
$(N_{\mbox{{\tiny obs}}} - N_{\mbox{{\tiny bkg}}})
/(a \times \epsilon \times BR(t\bar{t} \rightarrow b\ell^+\nu_\ell\bar{b}\ell^{\prime -}\bar{\nu}_{\ell^\prime}) \times \mbox{{\footnotesize \intlum}})$, 
are
$8.4^{+3.2}_{-2.7} \mbox{}^{+1.5}_{-1.1} \pm 0.5$~pb for the DIL and
$7.0^{+2.7}_{-2.3} \mbox{}^{+1.5}_{-1.3} \pm 0.4$~pb for the LTRK analysis, where
the first two uncertainties are statistical and systematic
and the third is due to the luminosity determination. 
We combine these results by dividing the analyses' expected
signal and background into three disjoint regions (DIL-only,
LTRK-only, and the overlap). Eleven events are shared
between DIL and LTRK.  Using the combined $a \times \epsilon \times BR(t\bar{t} \rightarrow b\ell^+\nu_\ell\bar{b}\ell^{\prime -}\bar{\nu}_{\ell^\prime})$ of 1.03\% and
accounting for common systematic uncertainties, a joint Poisson
likelihood is maximized yielding~\cite{comb}
$$
\sigma_{t\bar{t}} = 7.0^{+2.4}_{-2.1} \mbox{(stat.)} ^{+1.6}_{-1.1} \mbox{(syst.)} 
\pm 0.4 \mbox{(lum.)}~\mbox{pb}
$$

  We have performed several cross-checks.  
The techniques reproduce the expected $W$ and $Z$ production 
cross sections ({\it e.g.} $252 \pm 5$~pb measured vs. $252 \pm 0.9$~pb 
expected for LTRK $e$+track $Z$ candidates).
We compare the number of events with identified bottom quark jets 
in the signal sample to expectations and find agreement within uncertainties ({\it e.g.} 7 observed 
vs. $5.9 \pm 1.8$ expected for DIL). The measured \ttbar cross section
is stable within its uncertainty to variations of the loose and tight lepton $p_T$ and $E_T$ cuts.
When we restrict the analysis to two ``tight'' isolated
leptons, 
an expected signal-to-background ratio 
of 3.4 is achieved with $a \times \epsilon \times BR(t\bar{t} \rightarrow b\ell^+\nu_\ell\bar{b}\ell^{\prime -}\bar{\nu}_{\ell^\prime})$ = $(0.34\pm 0.05)$\%.   
We observe 7 candidates with a predicted background of $1.3\pm 0.5$ events, yielding
a cross section of  
$8.5^{+4.5}_{-3.5}\mbox{(stat.)} ^{+1.8}_{-1.4}\mbox{(syst.)}\pm 0.5\mbox{(lum.)}$~pb, 
in good agreement with the larger samples. 

\begin{figure}[h]
\includegraphics[width = 8.6cm]{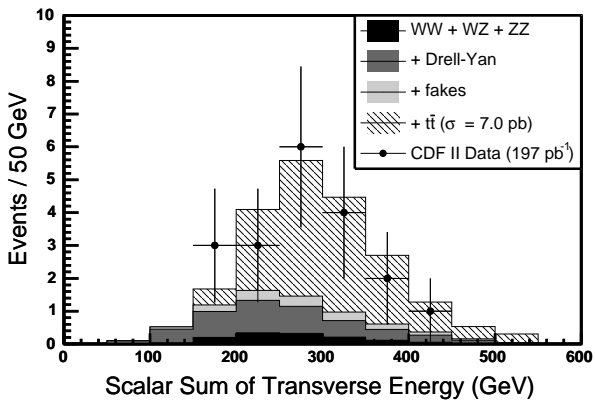}
\caption{\label{ht} 
$H_T$ (defined in text) for events from the LTRK
analysis with $\ge 2$ jets.}
\end{figure}

  We present key kinematical distributions of the signal sample and find good
agreement with the SM, assuming $m_{top}=175$~GeV/$c^2$.   
For example, using events from the LTRK analysis, Figure~\ref{ht}
shows a distribution of the previously defined $H_T$ variable.
A Kolmogorov-Smirnov test of this distribution yields a $p$-value of 75\%.

\begin{figure}[h]
\includegraphics[width = 8.6cm]{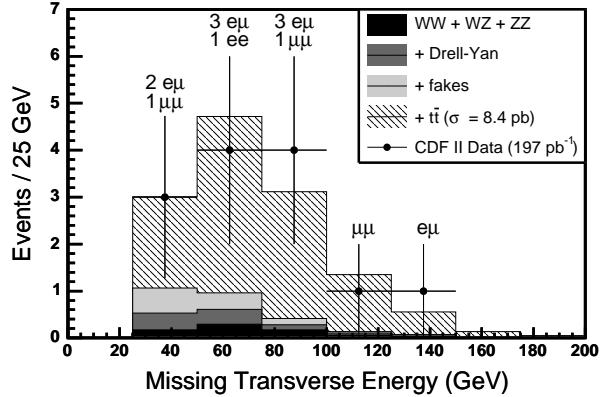}
\caption{\label{metdphi}
\MEt~for events from the DIL
analysis with $H_T > 200$ GeV and
$\ge 2$ jets.}
\end{figure}

In the DIL analysis, both leptons are always identified as
either an electron or a muon.  In Run I seven of the nine
observed events were e$\mu$, and these events populated the tail
of the expected \MET distribution.  The expected numbers of ee,
$\mu\mu$, and e$\mu$ events for the Run II DIL analysis, scaled
to the 13 total observed events, are $3.3 \pm 0.5$, $2.8 \pm 0.5$, and
$6.8 \pm 0.8$, respectively.  One ee, three $\mu\mu$, and nine
e$\mu$ events are observed in the data; the \MET for these
events is shown in Figure~\ref{metdphi}.
A Kolmogorov-Smirnov test of this distribution yields a $p$-value of 49\%.
%
%

  In summary, we have measured the \ttbar production 
cross section in the dilepton
channel to be $ 7.0^{+2.4}_{-2.1} \mbox{(stat.)} ^{+1.6}_{-1.1} \mbox{(syst.)} 
\pm 0.4 \mbox{(lum.)}~\mbox{pb}$ for $m_{top} = 175$~GeV/$c^2$~\cite{mtop}
using data from the first two years of running of the upgraded
Tevatron Collider and CDF~II detector. We observe good agreement
between the data and the SM prediction in event yield and
key kinematic distributions. The measured $t\bar{t}$ cross section
agrees well with the full NLO SM prediction 
of $6.7^{+0.7}_{-0.9}$~pb~\cite{topxsect1}.

  We thank the Fermilab staff and the technical staffs of the participating institutions for their vital contributions. This work was supported by the U.S. Department of Energy and National Science Foundation; the Italian Istituto Nazionale di Fisica Nucleare; the Ministry of Education, Culture, Sports, Science and Technology of Japan; the Natural Sciences and Engineering Research Council of Canada; the National Science Council of the Republic of China; the Swiss National Science Foundation; the A.P. Sloan Foundation; the Bundesministerium fuer Bildung und Forschung, Germany; the Korean Science and Engineering Foundation and the Korean Research Foundation; the Particle Physics and Astronomy Research Council and the Royal Society, UK; the Russian Foundation for Basic Research; the Comision Interministerial de Ciencia y Tecnologia, Spain; in part by the European Community's Human Potential Programme under contract HPRN-CT-20002, Probe for New Physics; by the Research Fund of Istanbul University Project No. 1755/21122001; and by the Research Corporation. 

\bibliography{dilepton_prl}

\begin{thebibliography}{20}
\bibitem[{geo()}]{geomnote}
\bibinfo{note}{We use a cylindrical coordinate system about the beam axis in
  which $\theta$ is the polar angle, $\phi$ is the azimuthal angle, and $\eta
  \equiv -\ln \tan (\theta /2)$. $E_T \equiv E \sin \theta$ and $p_T \equiv p
  \sin \theta$ where $E$ is energy measured by the calorimeter and $p$ is
  momentum measured by the spectrometer. \MET$\equiv -\sum_i E_T^i {\bf n}_i$,
  where ${\bf n}_i$ is the unit vector in the azimuthal plane that points from
  the beamline to the $i$th calorimeter tower.}
\bibitem[{etp()}]{etpt}
\bibinfo{note}{For muons, transverse momentum (not energy) is used.}
\bibitem[{com()}]{comb}
\bibinfo{note}{The individual cross sections (DIL-only, LTRK-only, and overlap)
  range from $\sim 4-10$~pb which aggregately yield the combined result of
  7.0~pb.}
\bibitem[{mto()}]{mtop}
\bibinfo{note}{The dependence of the measured cross-section on $m_{top}$ is
  -(+) 0.04 pb per GeV/c$^{2}$ greater(less) than the assumed value of 175
  GeV/c$^{2}$.}
\end{thebibliography}

\end{document}